\documentclass[epsfile,prc,aps,nofootinbib,showpacs]{revtex4}
\usepackage{graphicx}
\usepackage{epsfig}
\usepackage{amsmath}
\usepackage{amssymb}

\newcommand{\be}{\begin{equation}}
\newcommand{\ee}{\end{equation}}
\newcommand{\bqa}{\begin{eqnarray}}
\newcommand{\eqa}{\end{eqnarray}}

\begin{document}
\date{\today}

\title{Baryonium, a common ground for atomic and high energy physics\footnote{Contribution to EXA2014 Vienna
to appear in a special edition of Hyperfine Interactions.}}

\author{S. Wycech\footnote{e-mail: wycech@fuw.edu.pl}}
\affiliation{National Centre for  Nuclear Studies,
Warsaw, Poland}

\author{ J.P. Dedonder\footnote{e-mail: jean-pierre.dedonder@univ-paris-diderot.fr} and B. Loiseau\footnote{e-mail: loiseau@lpnhe.in2p3.fr}}
\affiliation{ Sorbonne Universit\'es, Universit\'e Pierre et Marie Curie, Sorbonne Paris Cit\'e,
Universit\'e Paris Diderot and IN2P3-CNRS UMR 7585,
Laboratoire de Physique Nucl\'eaire et de Hautes \'Energies,
4 place Jussieu, 75252 Paris, France
 }

\begin{abstract}
Indications of the  existence of  quasi-bound states in  the   $N {\bar N}$ system are presented.
Measurements by BES discovered a broad enhancement  close to  the $p {\bar p}$ threshold in the $S$ wave, isospin $0$ state formed in radiative decays of
$J/ \psi $. Another enhancement located about $50$~MeV below the threshold was found in mesonic decays of $J/ \psi $. In terms of the Paris potential  model
it was shown that these are likely to represent the same state.
Antiprotonic atomic data provide some support for this interpretation and indicate the existence of another fairly  narrow  quasi-bound state
in a $P$ wave.
\end{abstract}

\maketitle

\section{Introduction}
\label{intro}

 Nucleon-antinucleon quasi-bound states, or states coupled to these, were searched for   in the days of  LEAR at CERN.
 Nothing has been  found,
 but broad states or states close to the  threshold  were not excluded.  References~\cite{BER82, ADI86} indicate  conclusions
 of a long series of measurements.
In reference~\cite{ADI86} a search for\emph{ narrow} signals in the $\gamma$ spectrum from $p \bar{p}$ annihilation at rest
was performed and no discoveries  were found in the  region below $1770$~MeV  and $ \Gamma < 25$~MeV.
Experiments looking for missing mass in  reaction $p \overline{p}\rightarrow X \pi$ or $p \overline{p}\rightarrow X p$  brought
similar conclusions.
On the experimental side, one possible reason for  the failure is  the heavy
background due to annihilation processes. Another is the large number of allowed partial waves. On the theory side, it was assumed  that
the annihilation reaction involve   $\sim 2 M_p$ mass  transfer and by the uncertainty principle it  has to be very short ranged.
It was thus expected that widths of quasi-bound states might be narrow. The first part of the argument is still true but it is also
known  from scattering data that the annihilation potential is strong  already at $p \overline{p}$ separations of $ 1 fm $.

\begin{table}[ht]
\caption{ Low energy  $p \bar{p}$  states allowed in the $ J/\psi \rightarrow \gamma  p\bar{p}$ decays. The first column gives  decay  modes
and specifies the  internal states of $ p\bar{p} $ pair.  For both  photon and $ J/\psi $ the $J^{PC}  = 1^{(--)}$. The second column gives
$J^{PC}$ for the $p \bar{p}$ system. }

\begin{tabular}{|l|c|}
 \hline
decay mode                               &  $J^{PC}( p \bar{p} ) $       \\
\hline
$\gamma  p \bar{p} (^1S_0) $               &  $0^{-+}$                          \\
$\gamma  p \bar{p} (^3P_0) $               &  $ 0^{++}$                          \\
$\gamma  p \bar{p} (^3P_1)$                &  $ 1^{++}$                        \\
\hline
\end{tabular}
\label{table0}
\end{table}

A convincing detection requires  selective experiments, and the first such measurement   is the decay

\begin{equation}
\label{1}
 J/\psi \rightarrow \gamma p\overline{p},
\end{equation}
studied  by the BES Collaboration~\cite{BAI03}. A strong threshold
enhancement is observed in the invariant  $p \bar{p}$  mass distribution (see Fig.~\ref{fig-2gamdist}).
There are three  final $p \bar{p}$ states allowed by  $P$ and $C$
conservation in the $\gamma p \bar{p}$ channel. These are listed in
  tables~\ref{table0}    and~\ref{tableCP}
 and  denoted by  $^{2S+1}L_{J}$ or $^{2I+1,2S+1}L_{J}$, $S,L,J$
being the spin, angular momentum,  total momentum of the pair and
$I$ denotes the isospin. Radiative decay does not conserve isospin but already in Ref.~\cite{BAI03}
it was realized  that  $ I=0$ is the state  which leads to the enhancement.
From potential descriptions of $N \bar{N}$ interactions based on the $G$-parity rule it is known that the
 pion exchange potential is very strong in this state being capable to form bound states. On the other hand
 the annihilation and short range interactions act repulsively, thus  quasi-bound states are not guaranteed.
In particular  the Paris potential  generates a $52$~MeV broad
quasi-bound state at $4.8$~MeV below threshold~\cite{lac09} but the the Bonn-J\"ulich potential does not generate bound state in this
 wave~\cite{jul06}. Both can  describe the threshold enhancement~\cite{loi05, jul06}.

The radiative process  itself is puzzling  as the decay rate is comparable to the mesonic decay rates. On the other hand
the   conventional  coupling constants  $ \alpha / g^2_{NN,meson} $ are   $ \sim10^{-3}$ and a strong enhancement mechanism
 has to exist.  In  section~\ref{J}
this question is discussed  jointly with the origin of the threshold enhancement.
\begin{table}[ht]
\caption{ Decay modes and the   $p \bar{p}$  states allowed in the $ J/\psi \rightarrow ~ boson ~ p \bar{p}$  decays.   }
\begin{tabular}{|l|c|c|}
\hline decay mode                               &  branching   & $p \bar{p}$ states allowed      \\\hline
$\gamma  p \bar{p}  $            &  $3.8(\pm1.0)\cdot10^{-4}$ \cite{PDG10}  &  $^{1}S_0,  ^{3}P_1, ^{3}P_0$         \\
$\omega  p \bar{p} $             &  $1.1(\pm0.15)\cdot10^{-3}$\cite{abl07a} &  $^{11}S_0,  ^{13}P_1, ^{13}P_0$     \\
$ \pi^{0}  p \bar{p} $           &  $1.19(\pm.08)\cdot10^{-3}$\cite{PDG10}  & $^{33}S_1 , ^{31}P_1$                 \\
$ p \bar{p}  $                   &   $2.12(\pm0.1)\cdot10^{-3}$  \cite{PDG10} & $^{13}S_1 $     \\
\hline
\end{tabular} \label{tableCP}
\end{table}
 There are other indications of   $N\bar{p}$  structures existing below  threshold coming from antiprotonic atoms. These are discussed
 in  section~\ref{A}.
In section~\ref{newera} we list possible experimental researches of the baryonia which could be performed
in a near future.

\section{Final state interactions in $J/\psi$ decays}
\label{J}
We have attempted calculations of the radiative and mesonic decay rates presented in table \ref{tableCP}, assuming that mesons are emitted in the final states of the decay when the baryons have been formed. With conventional  meson-nucleon coupling constants this model reproduces
branching ratios of  $meson\ p \bar{p}$ channels relative to the basic  $ p \bar{p}  $  channel~\cite{DED14}.
It  offers also
a consistent description of the spectra in cases of $ \pi^0,\omega$  mesons at the
expense of one free parameter, the  $ N \bar{N}  $  formation radius  $R=0.28 ~ fm $.
However, this model fails in the description
of radiative decays in two ways: first the branching ratio is only about $1/3$ of the experimental one, and second  the threshold enhancement is not reproduced. The transition  to $^1S_0$ state is a magnetic one and it turns out that in the intermediate stage of such decay one has both
 $ p \bar{p}  $  and  $ n \bar{n}  $  as intermediate states. Since the magnetic moments of $p$ and $n$ have opposite sign the effect of
 final state enhancement  cancels strongly. At the same time such a model indicates that final state interactions increase the overall  decay rate by an
 order of magnitude in the  $I=1$ states.

\begin{figure}[ht]
\centering
\includegraphics[width=4cm,clip]{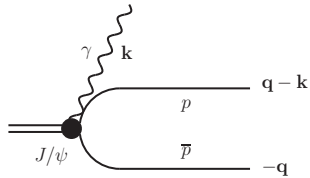}
\caption{The photon is  emitted either from $J/\psi$ or during the hadronisation   stage of the process and the final baryons are formed
in the $S$ wave}
\label{fig-1}      
\end{figure}

\begin{figure}[ht]
\centering
\includegraphics[width=5cm]{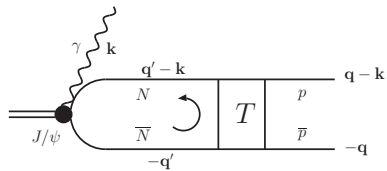}
\caption{The final state interaction is described by half-off shell  $T$ matrix
generated by the  Paris potential}
\label{fig-2}      
\end{figure}

\begin{figure}[ht]
\centering
\includegraphics[width=5cm, angle=90]{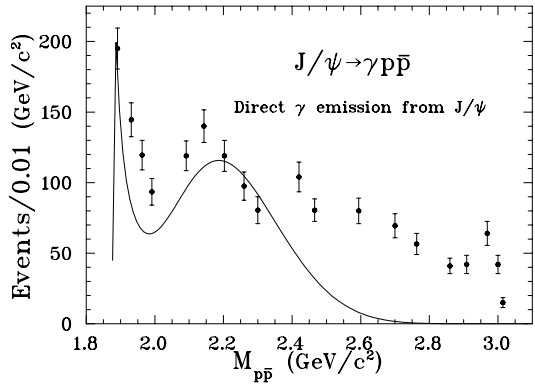}
\caption{The  $p \overline{p}$  invariant mass spectrum  obtained under the assumption that the photon is emitted  before the baryons are formed.
The missing strength at large $p\overline{p}$ invariant mass, $M_{p\overline{p}}$, comes from the photon radiated by final hadrons~\cite{DED14}
 }
\label{fig-2gamdist} 
\end{figure}

\begin{figure}[ht]
\centering
\includegraphics[width=5cm, angle=90]{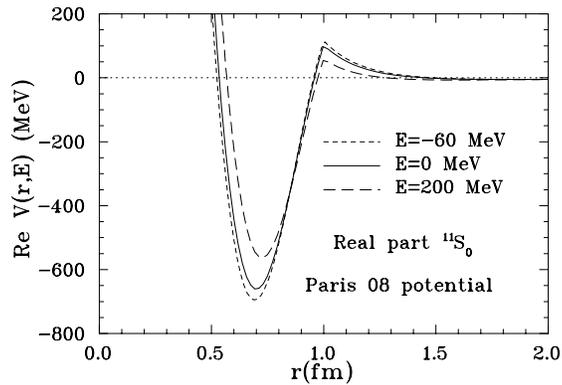}
\caption{The Paris $N \overline{N}$ real  potential in the $ ^{11}S_0$ wave. It generates a $50$~MeV broad quasi-bound state
at~$ \sim 5 $~MeV binding. The well and   barrier structure generate the  shape resonance visible in the spectrum
in figure~\ref{fig-2gamdist}  at~$ 2170 $~MeV}
\label{fig-4parispot}
\end{figure}

In this note  we report an extension of the FSI calculations  of Ref.~\cite{loi05} which  is now used to cover  the whole photon  spectrum. The basic
assumption of this approach  (also that in  Ref.~\cite{jul06}) is that the photon is emitted before the baryons are formed. The two related processes
are in the diagrams of Figs.~\ref{fig-1} and~\ref{fig-2} and the FSI is calculated in terms of half-off shell $T$ matrix
generated by the Paris potential~\cite{lac09} plotted in figure~\ref{fig-4parispot}.
This approach allows to calculate the spectrum but not the absolute decay rate. One free parameter, the radius $R$ (=0.28~fm)
of a  Gaussian source function is used to describe  the creation of a $\gamma p\overline{p}$  state (see figure~\ref{fig-1}). However, in order to reproduce  in a better way both maxima in figure \ref{fig-2gamdist} (the $ X(1859) $ and $X(2170)$ in BES terminology), it turns out profitable to assume the radius to be weakly  dependent  on the photon energy. It was found to  change from $0.28$~fm at maximal  $k\sim1.2$ GeV
to $0.39$~fm at  $k=0$.

Inspection of figure~\ref{fig-2gamdist} shows that both states may be reproduced by the $N \overline{N}$ potential related at
large distances via G-parity transformation to the $N N$ interactions. However,  the proper description of both peaks involves
distant extrapolation of the $T_{N\overline{N}}$ matrices off energy shell, which  corresponds to  very short range interactions.
This figure shows also that a
sizable portion of the spectrum is missing  and this  part  comes from  photon emissions by final  $N\overline{N}$  baryons and exchange currents
\cite{DED14}.

As already discussed the threshold enhancement indicates a "nearby singularity" that might describe an analogue of the bound state or the virtual
state known from the physics of two nucleons. To discern these possibilities  one has to test directly the sub-threshold region.

\section{Studies of  the $N \overline{N}$ sub-threshold region}
\label{A}
One  way to look below
the threshold is the detection  of $N \bar{N}$ decay products. The
specific decay mode
\begin{equation}
\label{3i}
 J/\psi \rightarrow  \gamma  \pi^+ \pi^- \eta'
\end{equation}
has been studied by the BES collaboration~\cite{abl05}.
 This reaction  is attributed by BES to an  intermediate
 $p \bar{p}$ configuration in the  $J^{PC}( p \bar{p} ) = 0^{-+}$ state
that is the $^1S_0$  wave.
A peak in the invariant  mass of the mesons was  observed and was
  interpreted as a new baryon state  and  named  X(1835).

\begin{figure}[ht]
\centering
\includegraphics[width=4cm,clip,angle=90]{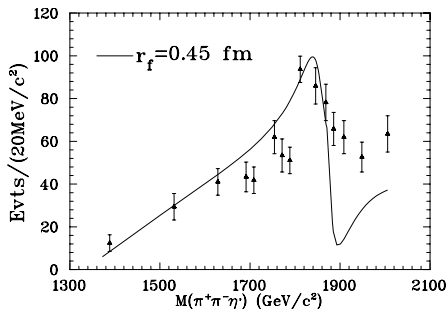}
 \caption{ The spectral function X$_S$ representing the $X(1835)$ shape.
 The parameter of the annihilation range  is $r_f=0.45~fm$.
This $S$-wave contribution has been normalized to reproduce the data close to the $X(1835)$ peak. The experimental points are from Ref.~\cite{abl05},
calculation from Ref.~\cite{ded09}. }
\label{fig-2XS}
\end{figure}

Under the assumption that all  mesons are produced in relative
$S$-waves the reaction (\ref{3i}), if attributed to an intermediate
$p \bar{p}$, is even more restrictive than the reaction (\ref{1}).
It allows only  one intermediate state, the
 $p\bar{p}$ $^{1}S_0$, which coincides with the previous findings.
The intermediate state of $p \bar{p}$ in reaction (\ref{3i}) is
possible but not warranted. In Ref.~\cite{ded09} a more consistent
interpretation is obtained with the dominance of
the $^{11}S_0$
state which  is a mixture of $p\bar{p}$ and $n\bar{n}$ pairs. It
has been  argued  that the peak  is due to an interference of a
quasi-bound, isospin 0,  $ N \bar{N}$ state with a background amplitude. A typical interference pattern obtained in this way
is plotted in figure~\ref{fig-2XS}. It is fairly close to the data. The same  quasi-bound
state was found in Ref.~\cite{loi05} to be  responsible for the threshold  enhancement in reaction (\ref{1}).
In this sense Paris potential unifies the two effects and attributes it to single quasi-bound state with an
energy dependent width.

Testing  the  subthreshold amplitudes  may be also realized in few body systems in particular in light antiprotonic atoms or at
extreme nuclear peripheries.
 In these conditions nucleons are bound and the effective subthreshold energies are  composed of binding energies and recoil  of the $N\overline{p}$
 pair  with respect to the rest of the system. For valence nucleons the   $ E_{binding}+ E_{recoil} $ may reach down to - 40~MeV below threshold.
 Let us indicate an  atomic   experiment that discovered an interesting anomaly.
Table~\ref{tab:radio} shows the ratios of antiproton capture rates  on neutrons and protons  $ C(n \overline{p})/ C(p \overline{p})$ bound to nuclear peripheries. These reflect the ratios of neutron and proton  densities. The second and third columns indicate such ratios extracted from   widths of two    antiprotonic
atomic  levels the "lower" and the "upper" one. These widths are determined at  nuclear densities   $ \sim 10 \% $ and $ \sim 5\%$ of the central density $\rho_0$. The last  column is obtained
with radiochemical studies of final nuclei with  one neutron or one proton removed in the annihilation reaction~\cite{TRZ03}.
The latter process is localized
at densities  $\rho \sim 10^{-3} \rho_0$.  In standard nuclei shown in the upper part of the table the ratios $ N(n \overline{p})/ N(p \overline{p}) \sim \rho_n /\rho_p $
increase at nuclear peripheries. However, in some nuclei characterized by small  proton  binding, indicated  in the lower part of the table,  and typical  ($\sim$ 8~MeV) neutron binding the ratio $ N(n \overline{p})/ N(p \overline{p})$ suddenly drops at extreme nuclear peripheries. That effect cannot be explained by the nuclear structure alone  and we attribute it to the
existence  of a narrow bound state in the $N \overline{N}$ system. Such a narrow state is in fact predicted by the Paris potential  in the  $^{33}P_1$  wave, see table~\ref{tab-1}.

\begin{table}[ht]
\centering
\caption{Ratios of  $ N(n \overline{p}) $ and $ N(p \overline{p}) $ capture
rates from atomic states. The last  column  shows experimental
numbers from radiochemical experiments. Other columns (see text) give ratios
calculated with optical potential and plausible nuclear
densities  based on experimental results
from Ref.~\cite{TRZ03}.}
 \label{tab:radio}
  \begin{tabular}{llll}
  \hline
atom & lower & upper& radiochemistry  \\\hline
 $^{96}$Zr &0.95(9)&1.53(29)&2.6(3) \\
$^{124}$Sn &1.79(10)&2.44(39)&5.0(6) \\ \hline
$^{106}$Cd &1.64(80)&2.10(80)&0.5(1) \\
$^{112}$Sn &1.90(13)&2.43(49)&0.79(14)\\
\end{tabular}
\end{table}
\begin{table}[ht]
\centering
\caption{Binding energies  in MeV of the close to
threshold quasi-bound states in  the Paris potential ~\cite{lac09} }
\label{tab-1}
\begin{tabular}{ll}
\hline
$^{2T+1\ 2S+1}L_J$ &  $ E - i \Gamma/2$   \\\hline
$^{11}S_0$ & -4.8-i26  \\
$^{33}P_1$ & -4.5-i9.0  \\\hline
\end{tabular}
\end{table}

\begin{figure}
\includegraphics*[height=7.5cm]{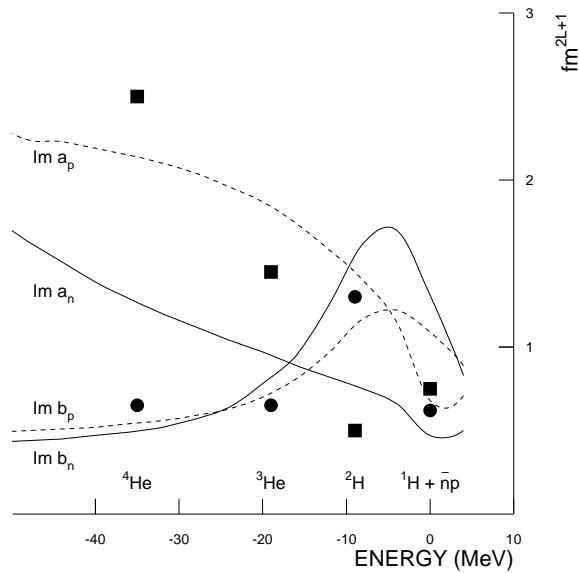}
 \caption{\label{fig5} The absorptive parts  of spin-isospin
averaged $N \bar{p} $ scattering amplitudes extracted
 from the atomic level widths in antprotonic H, $^2$H,
$^3$He and $^4$He~\cite{lightatoms1, lightatoms2}. Squares: $S$ waves and
circles: $P$ waves. The bottom scale indicates the energy below
threshold.
 The curves, calculated with the
 Paris   potential, give the
amplitudes separately:  $a_{n(p)}$ denote the $n \bar{p} $ or $p \bar{p} $ $S$-wave amplitudes. Similarly  $b_{n(p)}$,  $b_{p(p)}$
are the corresponding $P$-wave  amplitudes. The strong increase of absorption in the
$\bar{p}p$ $S$  wave is attributed mainly to the $^{11}S_0 $
state. }
\end{figure}

A similar effect is indicated by studies of experimental absorption lengths in light antiprotonic 
atoms~\cite{lightatoms1, lightatoms2}.  From the lower and upper
atomic level widths one can extract average  $S$ wave absorption length  Im~$a$     and $ P$ wave absorption volumes Im~$b$.
The results shown in figure~\ref{fig5} indicate increase of the absorption in the $S$ wave down below the threshold. This result is
consistent with the presence of the $ X(1835)$ state.   The absorption volume extracted from antiprotonic deuterium indicates some enhancement
possibly related to radiochemical anomalies observed in nuclei  with loosely bound protons and interpreted in terms of Paris potential as the  $ P$ wave
bound state.

\section{New era}
\label{newera}
Following indications  from antiprotonic atoms and from BES experiments  the  baryonia should be searched in the region of 0-60~MeV below  the $N \overline{N}$ threshold.
With the new $\bar p$ beams expected to operate in J-PARC and FAIR it would be advisable to repeat two old experiments possibly at  different energies, possibly with polarized particles:

$\bullet$ Search for narrow signals in the $\gamma$-spectrum from $p \overline{p}$ annihilation was performed  at rest~\cite{ADI86},
The signals (in the region that we  expect them now to exist) were  covered by heavy background due to $\pi^0$ decays and
$\pi^-p \rightarrow \gamma n $. It would be better to perform this experiment with higher energy antiprotons  (a few hundred~MeV/c)
 antiprotons which could shift the expected signal away from  the  heavy  background region.

$\bullet $  The  $ \overline{p} d \rightarrow n  X $ experiment~\cite{BER82} was performed at $ 1.3 $~GeV/c. This gives rather  small
chance of $p\overline{p}$ coupling in the statistically insignificant $ ^{11}S_0$ wave. Lower energies and polarized (one or two particles)
 would reduce the background.

New instructive experiments that possibly could be performed at FLAIR are:

$\bullet$  Fine structure splitting in light antiprotonic atoms $^1H,^2H,^3H,^3He,^4He$ would allow to trace energy dependence
of the selected $ \overline{p}N$ amplitudes in the subthreshold region down to $ \sim - 40 $~MeV.

$\bullet$  Studies of mesons emitted  from annihilations of $ \overline{p} $  at nuclear peripheries. In particular nuclei with closed shells
 with one loosely bound valence nucleon could be profitable. In the latter case the baryonium  signal would be separated from
 a complicated background due to  other annihilation channels.

\vspace{0.2cm}
\noindent
This work has been partially supported by a grant from the French-Polish exchange program COPIN/CNRS-
IN2P3, collaboration 05-115.
SW was also supported by  Narodowe Centrum Nauki  grant 2011/03/B/ST2/00270.

\end{document}